\documentclass[12pt]{article}

\usepackage{latexsym}
\usepackage{amssymb}

\usepackage{amsmath}
\usepackage{amsfonts}
\usepackage{epsfig}

\newtheorem{theo}{Theorem}

\newtheorem{lemm}{Lemma}

\def\NNN{{\rm I\kern-.2em N}}          
\def\N{{\cal N}}          
          
\def\tA{\tilde{A}}          
\def\tsigma{\tilde{\sigma}}          
          
\def\E{\mathbb{E}}
\def\S{{\cal S}}

\def\Z{\mathbb{Z}}
\def\R{\mathbb{R}}
\def\B{{\cal B}}

\def\G{{\cal G}}

\renewcommand{\E}{\mathbb E \,}

\newcommand{\RR}{{\cal B}}
\newcommand{\rr}{{\cal R}}

\newcommand{\tod}{\stackrel{{\cal D}}{\longrightarrow}}
\newcommand{\D}{{\cal D}}

\newcommand{\toL}{\stackrel{{L^1}}{\longrightarrow}}

\newcommand{\toLp}{\stackrel{{L^p}}{\longrightarrow}}
\newcommand{\lb}{\left(}
\newcommand{\rb}{\right)}
\newcommand{\eqco}{\setcounter{equation}{0}}
\newcommand{\thco}{\setcounter{theo}{0}}
\newcommand{\prco}{\setcounter{prop}{0}}
\newcommand{\laco}{\setcounter{lemm}{0}}
\newcommand{\coco}{\setcounter{coro}{0}}
\newcommand{\cjco}{\setcounter{conj}{0}}

\newcommand{\deco}{\setcounter{defn}{0}}
\newcommand{\reco}{\setcounter{rema}{0}}
\newcommand{\allco}{\eqco  \thco \prco \laco \coco \cjco \deco \reco}
\newcommand{\qed}{\rule[-1mm]{3mm}{3mm}}

\newcommand{\Po}{{\cal P}}
\newcommand{\Q}{{\cal Q}}

\newcommand{\Var}{{\rm Var}}
\newcommand{\var}{{\rm Var}}

\newcommand{\X}{{\cal X}}
\newcommand{\A}{{\cal A}}
\newcommand{\F}{{\cal F}}

\newcommand{\NN}{{\cal N}}

\newcommand{\eps}{\varepsilon}
\def\bdm{\begin{displaymath}}
\newcommand{\edm}{\end{displaymath}}
\def\benu{\begin{enumerate}}
\def\eenu{\end{enumerate}}
\def\beqn{\begin{equation}}
\def\eeqn{\end{equation}}
\def\be{\begin{equation}}
\def\ee{\end{equation}}
\def\bea{\begin{eqnarray}}
\def\eea{\end{eqnarray}}
\newcommand{\bean}{\begin{eqnarray*}}
\newcommand{\eean}{\end{eqnarray*}}
\newcommand{\bear}{\begin{eqnarray}}
\newcommand{\eear}{\end{eqnarray}}

\newcommand{\0}{{\bf 0}}

\renewcommand{\epsilon}{\varepsilon}

\begin{document}

\title {\bf Limit theorems for monolayer ballistic deposition in
the continuum } 
\bigskip
\author{Mathew D. Penrose$^{1}$}
\date{May 2001}
\maketitle

\begin{abstract} 
We consider a deposition model in which  balls
rain down at random towards a 2-dimensional surface, roll
downwards over existing adsorbed balls, are
adsorbed if they reach the surface, and discarded 
if not.  We  prove a spatial law of large numbers
and central limit theorem  for the ultimate number
of balls adsorbed onto a large toroidal surface,
and  also
for the number of balls adsorbed on the restriction to a large region
of an infinite surface.

\end{abstract}


\footnotetext{$^1$Department of Mathematical Sciences,
University of Durham, South Road, Durham DH1 3LE, England:
{\texttt mathew.penrose@durham.ac.uk}, (44)191 374 384, fax (44)191 374 7388}

\section{ Introduction and statement of results}
\allco

{\em Random sequential adsorption} (RSA) is a mathematical model,
 incorporating stochastic and geometric elements,
 for sequential deposition of colloidal particles
or  proteins onto a surface (or {\em substrate});
 particles  arrive at random locations,
and each adsorbed  particle occupies a region of the substrate which
prevents the adsorption of any subsequently arriving
particle in an overlapping surface region. 
 Scientific interest is considerable; for a series of surveys, see
{\em Colloids and Surfaces A}, Volume 165 (2000), for example 
 Privman \cite{Pr},
 Senger {\em et al.} \cite{Senger}, Talbot
 {\em et al.} \cite{Talbot} and Wang \cite{Wa}.
  See Evans \cite{E} for a much-cited earlier survey.

For deposition onto a surface of dimension $d$,
there have been many simulation studies, often concerned 
with the number of particles
 ultimately adsorbed onto a  region of substrate.
It is of interest to know whether
 this satisfies a law of large numbers (LLN, i.e. a thermodynamic limit) and
central limit theorem (CLT, i.e. Gaussian fluctuations)
 as the region becomes large.
While previous rigorous mathematical studies were mainly restricted
to 1 dimension, for general $d$
Penrose \cite{Plln} proved a LLN for various
continuum systems, and  \cite{Pinf} proved both LLNs  and CLTs
 for certain lattice systems.
 Penrose  and Yukich \cite{PY2} proved both LLNs and CLTs for continuum systems 
with {\em finite input}  where
the addition of incoming particles
is  terminated before saturation occurs.
Except in the case $d=1$ (Dvoretzky and Robbins \cite{DR}),
a CLT for infinite-input continuum  RSA remains
elusive.

In the present work we prove 
a  LLN and CLT for an infinite-input continuum
model related to RSA which has received 
attention in its own right on grounds of realism, namely,
 a form of monolayer {\em ballistic deposition} (BD), 
representing deposition in the presence of a gravitational field.
 Each incoming particle occupies a Euclidean  ball
of  radius $\rho$  in  $\R^{d+1}$, 
with  $d=2$ or $d=1$;
 the $(d+1)$-st coordinate  represents  ``height''.
An incoming particle
falls perpendicularly from above towards  a substrate 
represented by the
 surface $\R^d \times \{0\} \subset \R^{d+1}$, which
we identify with the lower-dimensional space $\R^d$,
or a sub-region thereof (the {\em target region}).
Its downward motion  is vertical until it hits
the substrate or one of the particles previously adsorbed.
If  it contacts a
previously deposited particle, then the new particle  rolls, following
the path of steepest descent until it reaches a stable position.
If the new particle reaches the adsorption surface, it is fixed there;
otherwise it is removed from the system.  
For $d=1$, the  model dates back to Solomon (\cite{Solomon}, page 129),
and the  formulation in $d=2$ by  Jullien and Meakin \cite{JM}
has led to considerable renewed interest;  see  \cite{Senger,Talbot}.
We state and prove results only for $d=2$;  changing to $d=1$ 
makes things easier.

 To avoid having to specify  
 the  behaviour of particles near the boundary of the target region,
we assume, as in most simulation studies,
that the target region is a {\em torus} with
integer dimensions.  
Given $A \subset \Z^2$ of the form 
$A = \{m_1,\ldots, m_2 \} \times \{m_3,\ldots, m_4\}$
(a {\em lattice rectangle}),
define ${\tilde A} \subset \R^2$ by
\bea
\tA = (m_1-1,m_2 ]\times (m_3-1,m_4].
\label{tAdef}
\eea
We focus attention mainly on target  regions of this form,
and adopt periodic (toroidal) boundary conditions for the
 rolling mechanism.

Suppose $X_1,X_2, X_3,\ldots$ are independent and uniformly distributed
 over $\tilde{A}$. These form the random input 
to the model with target region $\tA$;
 the vector   $X_i$ represents the position at
 which the $i$-th incoming ball would end up touching
the $2$-dimensional substrate if it were
to fall un-hindered.
 Successive balls
are adsorbed (with possible displacement
due to rolling) or rejected  according
to the BD mechanism  described above, but
adopting the toroidal boundary conditions, whereby
 an adsorbed particle near one edge of the 
rectangle $\tilde{A}$ can influence
the rolling of a particle near the opposite edge.
The process terminates when there is  no available space left 
on the substrate large enough 
to contain a new item (jamming of $\tilde{A}$);
that is, when every point of $\tA$ lies within 
 a distance less than $2\rho$, using the toroidal metric,  from
 some point in $\tA$ that is the location of the point of
contact  of some previously adsorbed ball.

Let $N(A)$ denote the (random) number of balls adsorbed   at the
termination time.  Our first result is 
 a LLN for $N(A)$ as $A$ becomes large.
For any sequence of sets $(A_n)_{n \geq 1}$,
set
$$
\liminf(A_n) : = \cup_{n \geq 1}  \cap_{m \geq n } A_m.
$$
  For $p \geq 1$,
let $\toLp$ denote convergence in $p$-th moment
 as $n \to \infty$.

\begin{theo}
\label{th2}
There is a  constant $\mu = \mu(\rho) >0$
 such that if $(A_n)_{n \geq 1}$ is a sequence of
lattice rectangles 
 satisfying $\liminf(A_n) =\Z^d$,
 then for any $p \in [1,\infty)$,
\bea
\frac{ N(A_n) }{|A_n|} \toLp \mu.
\label{eqth2}
\eea 
\end{theo}

Next, we give an associated CLT.
Let $\N(0,\sigma^2)$ denote a normally distributed
   random variable with mean zero and variance $\sigma^2$, if
 $\sigma >0$, or a degenerate random variable taking the value 0 with
 probability 1, if $\sigma =0$.  Let $\tod$ denote convergence in distribution.

\begin{theo}
\label{clt2}
There is a constant $\sigma_1 = \sigma_1(\rho)  > 0$ such that
for any sequence $(A_n)_{n \geq 1}$
of lattice rectangles with $\liminf(A_n)= \Z^d$,
 we have as $n \to \infty$ that
\bea
|A_n|^{-1} \Var(N(A_n)) \to \sigma_1^2
\label{varcon2}
\eea
 and
$$
|A_n|^{-1/2} \lb N(A_n) - \E N(A_n))  \rb \tod \
\N(0,\sigma_1^2).
$$ 
\end{theo}

Various alternative boundary conditions,
other than the toroidal scheme above, are also feasible.
Particles could simply roll until they
touch the surface $\R^2$  (possibly outside the target
region); or any particle that 
ends up  touching the surface outside the target region could be removed;  
 or  (as in Solomon's \cite{Solomon} version
of this model, generalized to $d=2$ in Weiner \cite{Weiner1})
 the boundary itself could cause a deflection
of particles (imagine a `wall' around the boundary of the target
region). For the LLN, these boundary conditions are not so important,
and a result like Theorem \ref{th2} can be obtained any
boundary conditions provided the influence of the boundary
has finite range. Moreover, 
the target regions in the sequence do
 not need to be rectangular, provided
only that they satisfy a condition of vanishing boundary length
 relative to their area.
We do not go into details on such generalizations
because of their proximity to results in \cite{Plln}.
 For the CLT, however,
 alternative  boundary conditions can cause extra difficulties
in the proof.  We believe these can be overcome in at least some cases
of   non-toroidal boundary conditions, but have
not written out  the details.

While toroidal boundary conditions are usually
used in simulation studies, they are not so realistic physically.
Another way to avoid  boundary effects is to take the whole
 of $\R^2$ as target region.  Our next result  shows  that a
 stationary point process, 
loosely speaking the set of locations of adsorbed points for 
the BD process with target region $\R^2$,  exists as 
a weak limit of point processes arising from bounded  target regions.
  Let $\S$ be the space of locally finite subsets  of $\R^2$.
For  $\zeta \in \S$ and $B \subset \R^2$, let $\zeta (B)$ denote
the number of elements of $\zeta$ in $B$ (so $\zeta(\cdot)$ is 
  a counting measure).
A {\em point process} on $\R^2$ is a random element $\zeta$ of $\S$.
For more details,
see for example \cite{DVJ,SKM,Res}.

If $\zeta$ and $\zeta_n$ ($n \in \NNN$) are point processes on $\R^2$,
we say the sequence $\zeta_n$ converges weakly to $\zeta$
 if the finite-dimensional distributions
converge, i.e. if for any finite collection of bounded Borel sets $B_i$
satisfying $\zeta(\partial B_i)=0$ almost surely, the joint
probability  distributions
of ${\zeta_n}(B_i)$ converge weakly to those of
$\zeta(B_i)$. This is equivalent
to various other definitions of weak convergence; see
 e.g.  section 9.1 of \cite{DVJ}.

Given a lattice rectangle $A$, let $\xi^A$ be the point process
of locations in $\R^2$  of ultimately 
accepted particles, for the BD model with target region 
$\tilde{A}$ (which will be a point process in $\R^2$, all of
whose points lie in $\tA$).  Our next result concerns weak
convergence of the point process $\xi^A$ as $A$ becomes large. 
As with Theorem \ref{th2}, the result is not sensitive
to the toroidal boundary conditions.
 
\begin{theo}
\label{th2c}
There exists a  stationary point process $\xi$, such that
if $(A_n)_{n \geq 1}$ is any sequence of lattice rectangles
 with $\liminf_{n \to \infty} A_n =  \Z^2$, 
 the sequence of point processes 
$\xi^{A_n}$ converges weakly to $\xi$.
\end{theo}

Given any  bounded region $B\subset \R^2$,
the interpretation of $\xi(B)$ is as follows.
The variable $\xi^A(B)$  is the number of adsorbed
points in $B$ when the target region is $\tA$.
As $A$ becomes large this has a weak limit which
is $\xi(B)$. If one now, in turn, makes $B$ large,
it is of interest to know if $\xi(B)$  
satisfies a CLT, and our final  result says that this is
indeed the case. We restrict attention to rectangular
regions although other shaped regions can also be
dealt with (see the proof).
\begin{theo}
\label{cltII}
There exists a constant $\sigma_2= \sigma_2(\rho) > 0$ such that  
for any sequence $(B_n)_{n \geq 1}$
 of lattice rectangles with $\liminf(B_n)= \Z^d$,
 we have as $n \to \infty$ that
\bea
|B_n|^{-1} \Var(\xi(\tilde{B}_n)) \to \sigma_2^2
\label{varcon3}
\eea
 and
$$
|B_n|^{-1/2} \lb \xi(\tilde{B}_n) - \E \xi(\tilde{B}_n)  \rb \tod \
\N(0,\sigma_2^2).
$$ 
\end{theo}
Other properties of the point process $\xi$ are also of interest.
The proof of Theorem \ref{cltII} involves showing
that $\xi$ has exponentially decaying correlations,
which is of interest in its own right.

Weiner \cite{Weiner2} considered an alternative version of
the BD model in two (or more) dimensions in which the region of
 substrate occupied by a particle is a rectilinear square rather than a circle
(the `Solomon model').  He claimed to prove a CLT for the number
 of particles ultimately adsorbed onto a large target region. However,
 his argument uses assertions from Weiner \cite{Weiner1}, which 
he later retracted (Weiner \cite{Weiner3}).  
It is   possible to adapt
our methods  to yield
 a CLT for Weiner's `Solomon model', at least in
the case of toroidal boundary
conditions, partially vindicating 
Weiner's claims regarding this model (though not the `Renyi model').


\section{Geometric preliminaries}
\label{secpal}
\allco

In notation  used throughout this paper,
 $\0$ denotes the origin $(0,0)$ of $\R^2$.
For $x =(x_1,x_2) \in \R^2$, 
 $\|x\| $ denotes the Euclidean  norm (modulus)
$\sqrt{x_1^2+ x_2^2}$ of
$x$.
For $x \in \R^2$ and $R \subset \R^2$,  $x+R$ denotes the translated set 
$\{x+y: y \in R\}$.
For $r >0$, define the continuum disk and
 $D(x;r) \subset \R^2$
the  lattice ball $B(x;r)\subset \Z^2$ by 
$$
D(x;r) = \{y: \|y -x\| \leq r \}; ~~~~ B(x;r) = D(x;r)\cap \Z^2.
$$
If $E$ is an event in a given probability space let
${\bf 1}_E$  be the indicator random variable taking the value 1 if
$E$ occurs and 0 if not.
Finally, for any directed graph, by a {\em root} of
the directed graph we mean a vertex with indegree zero.

We start with
two purely geometric results
about the mechanics of the BD model with target region given
 by the infinite surface $\R^2 \times \{0\}$. 
Each accepted 
 particle lies on the substrate, and so can be represented by
the point in  $\R^2$ at which it touches the surface.
 The position of an accepted
particle is a translate (or {\em displacement}) of
 the  location in $\R^2$ above which it originally comes in.
 The displacement (and also the decision on whether or not to accept)
is determined by the initial location at which
the particle comes in, and the positions
(after displacement)
of the previously accepted particles.

\begin{lemm}
\label{tennislem}
With probability 1, no particle receives
a displacement of  modulus greater than $8 \rho$.
\end{lemm}

{\em Proof.} 
 Choi {\em et al.} \cite{CTTV}
enumerate the possible fates an incoming ball might undergo.
Since these involve at most 4 deflections, in effect they state
the result but do not give a complete proof. Therefore we 
do so here.

Let each particle already
accepted be  represented by a point in $\R^2$
located at its point of contact with the substrate.
For any two such points the inter-point distance
$r$ (say) satisfies $r\geq 2\rho$.

For a new particle, let it too be
represented by a point in $\R^2$, obtained
by projecting the position  of its center down
onto the substrate (imagine looking down
on the substrate from above). As it rolls,
the point representing the  new particle performs a piecewise
linear motion in $\R^2$. The first
 line segment of this motion
represents an initial period when the new particle
touches a single existing  particle, and is of length at
most $2 \rho$.
Thereafter, each successive line segment
will represent motion while touching two
existing particles, and will be along
the mediator (perpendicular bisector)
between the two points representing
those particles. Let $d_j$ denote the
distance between the two particles  which
the new particle touches during the
$j$-th step of linear motion,
and note that $2\rho \leq d_j < 4\rho$.

 Each change in direction
of this piecewise linear motion in $\R^2$,
 say from step $j$ to step $j+1$ of linear motion,
will occur at the circumcenter of three
existing points. If this circumcenter
lies inside the triangle with corners at
those three points, then the motion comes
to a stop and the particle is discarded,
according to the BD rules. Therefore
for the motion to continue,
 the circumcenter lies outside
this triangle, so that the triangle must
have an obtuse angle. The  inter-point distance
$d_{j+1}$ is the longest edge length of this
triangle, while $d_j$ is one of the other
two edge lengths.
Since the triangle  has an obtuse angle, and
 all three edges are of length at least $2\rho$,
we obtain
$$
d_{j+1}^2 \geq  d_j^2 +  4\rho^2 ,
$$
and since the  edge-lengths $d_j$  must all be at most  $4\rho$, this
means that the sequence $(d_j)$ terminates in at most three steps,
in addition to the initial rolling in contact with just a single
previous particle. 
Since each piecewise linear step is of length at most $2 \rho$, this
completes the proof. \qed \\

Let us say that  after $k$ adsorptions, a given point
$ x \in\R^2$ is  {\em available}
for a particle to  be adsorbed, if  there are no adsorbed particles
touching the surface in $D(x;2\rho)$.
\begin{lemm}
\label{bdlem1}
There exist $\eps_0>0, \eps_1 >0$ with the following
property.  
Suppose that for some $k$,  after $k$ adsorptions, a given point
$ x \in\R^2$ is  available for a particle to  be adsorbed.
  Then there exists  a region of area at least
$\eps_1$ such that any incoming particle with location
in that region will be adsorbed in a position that  makes all points in
 $D(x;\eps_0)$  unavailable.
\end{lemm}
{\em Proof.}
Assume without loss of generality
that $\rho = 1/2$ and $x = \0$.
Take $\eps_0 < 1/8$.
First  suppose no adsorbed particle lies
within distance $1+ \eps_0$
of the origin. Then any particle arriving
in the ball $D(\0;\eps_0)$ 
will be accepted without rolling, and
for such a new particle  the 
unit diameter ball centred at that particle covers
the ball $D(\0; \eps_0)$. 

Next suppose there already exists a particle 
(at $x$, say) with $1 \leq \|x\| < 1 + \eps_0$.
A particle arriving in $D(\0;\eps_0) \cap D(x;1)$
will receive a first deflection and
 roll, but not very far. This is because
its initial distance from $x$ is more than $1 -\eps_0$
so it initially rolls at most a distance $\eps_0$ before
it reaches the surface or receives  a
 second deflection.  If a second deflection takes place, at that
instant
the new particle  then lies on the mediator of two adsorbed points $x,y$ say. 
The distance between $x$ and $y$ is more than
 $2 (1- \eps_0)$. If a third deflection were to take
place it would be at the circumcenter of adsorbed points
points $x,y,z$, say, making a triangle with an obtuse angle. But this
is impossible; for example if $(y,z)$  were the longest edge,
the cosine rule would give us
$$
\|y-z\|^2  \geq \|x-z\|^2 + \|x-y\|^2 \geq 1 + (4 - 8 \eps_0  + 4 \eps_0^2),
$$  
and therefore $\|y-z\|>2$ and the third deflection does not take place.
Therefore after the second deflection the linear motion terminates
either with adsorption or rejection. By Pythagoras' theorem,
 the distance
travelled in this last linear  motion after the second deflection
is at most
$$
\sqrt{ 1   - (1 - \eps_0)^2 } \leq (2 \eps_0)^{1/2}
$$
and therefore the total displacement of the  particle
before  adsorption is at most $\eps_0 + (2 \eps_0)^{1/2}$.

Therefore  if a particle arrives within a distance $\eps_0$ 
of the origin, it is adsorbed or rejected  at 
a distance  at most $2 \eps_0 + (2 \eps_0)^{1/2}$ from the origin.
Since $\eps_0 < 1/8$, this  is at most $ 3/4$. If adsorbed,
it therefore prevents any subsequent adsorption taking place
in $D(\0;\eps_0)$.  Therefore we are done, unless
there is  a possibility of rejection for 
particles arriving in $D(\0;\eps_0)$.

Next, suppose that it is possible for a particle
arriving within distance $\eps_0$ of the origin to be
rejected. If this happens it will be at
the circumcenter of points $x,y,z$ (say) after initial
deflection by $x$ and subsequent deflection by $y$ (say).
In this case the circumcenter of $x,y,z$   is at a distance
less than 1 from each of $x,y,z$, and {\em every point
inside the triangle $xyz$ is unavailable.} 
In particular, the origin does not lie inside the triangle
$xyz$; however, it does lie within distance $2 \eps_0$  of
the midpoint of $x$ and $y$.

Now suppose a particle lands in $D(\0;\eps_0)$ 
on the same side of the line $xy$ as the origin.
In the course of its subsequent rolling it stays
on the same side of the line $xy$ as the origin;
if not there would be some other line segment
between adsorbed centres, other than $\{x,y\}$,
of length between $2(1-\eps_0)$ and $2$, whose midpoint
lies in $D(\0;2 \eps_0)$, which is geometrically impossible
provided $\eps_0$ is sufficiently small.
At the end of its motion, if it were rejected,
 that would take place  at the circumcenter of points
$xyz'$, say, and in that case  all points landing in $xyz'$ would
be unavailable. However, provided $\eps_0$ is small enough,
 the origin must lie in
$xyz'$, and therefore we would have a contradiction. 

It follows that provided
$\eps_0$ is small enough, a point landing in $D(\0;\eps_0)$ 
on the same side of the line $xy$
as the origin will be accepted, in a position
that makes the region $D(\0;\eps_0)$ unavailable.
The desired conclusion follows, with $\eps_1 = \pi \eps_0^2/2$. 
$\qed$

\section{Probabilistic preliminaries}
\label{secprob}
\allco

The author \cite{P,Pinf} has developed general LLNs and CLTs for
functionals on the restriction of spatial white noise
processes to finite regions of the lattice, as follows.
Suppose $(E,{\cal E},P_0)$ is  an arbitrary probability space,
and $X = (X_x, x \in \Z^2)$ is  a family of independent
identically distributed random elements of $E$, each $X_x$ having
distribution $P_0$.
Let $X'$ be the process $X$ with the value $X_\0$ at the origin
replaced by an independent copy $X_*$ of $X_\0$ (that is, an
$E$-valued variable $X_*$ with distribution $P_0$, independent of
$X$), but with the values at all other sites the same.
 Let $\rr$ denote some collection of  nonempty finite subsets 
of $\Z^2$, with $x+B \in \rr$ for all $B \in \rr, x \in \Z^2$.

A {\em stationary  $\rr$-indexed functional of $X$} is
 a family $H=(H(X;A), A \in \rr)$ of real-valued
 random variables,
with the property that $(X_x,x \in A)$ determines the value of
$H(X;A)$ and 
 $H(\tau_yX;y +A) = H(X;A)$ (almost surely) for all $y \in \Z^2$,
   where $\tau_yX$ is the family of variables $(X_{x-y},x \in \Z^2)$.
Let $\Delta_0(A)$ be the increment $H(X;A) -H(X';A)$.  The  functional $H$
 {\em stabilizes on sequences tending to $\Z^2$}
  if there exists a  random variable $ \Delta_0(\infty)$
 such that for any  $\rr$-valued sequence $(A_n)_{n \geq 1}$ with
$\liminf_{n \to \infty}(A_n )= \Z^d$, the variables 
$\Delta_0(A_n)$ converge in probability to $\Delta_0(\infty)$.

A {\em stationary $\rr$-indexed summand}
is a collection  $(Y_{z}(X;A),  A \in \rr,z \in A)$ 
of real-valued random variables with the property
that  $ (X_x,x \in A)$ determines the value of
$Y_{z}(X;A)$, and $Y_{y+z}(\tau_y X;y+A) = Y_{z}(X;A)$ (almost surely) 
for all $y \in \Z^d, A \in \rr, z \in A$.
The associated {\em  induced stationary $\rr$-indexed sum}
is given by $H(X;A)= \sum_{z \in A} Y_{z}(X;A)$,
 which  is a   stationary $\rr$-indexed functional.

We restrict attention  here to the case where
$\rr$ is the collection,  here  denoted  $\RR$, of
 all lattice rectangles $\{m_1,\ldots,m_2\}\times \{m_3,\ldots,m_4\}$
with $m_2 -m_1 > 20 \rho$ and $m_4 -m_3 > 20\rho$.
This is different from the class of sets
denoted  $\RR$ in \cite{Pinf}; nevertheless,
the following general law of large numbers  is  
proved in just the same manner as the first part of
Theorem 3.1  of \cite{Pinf}.

\begin{lemm}
\label{genllnclt}
Suppose $(Y_{z}(X;A):  A \in \RR, z \in A)$ is a stationary $\RR$-indexed
summand inducing a stationary $\RR$-indexed sum $(H(X;A); A \in \RR)$.
Suppose that
\bea
\sup \{ \E |Y_\0(X;A)|:  A \in \RR: \0 \in A\} < \infty.
\label{unimom}
\eea
Suppose there exists an integrable random variable $Y_\0(X)$ such that
$Y_{\0}(X;B_n) \to Y_\0(X)$ in $L^1$ as $n \to \infty$,
for any $\RR$-valued sequence $(B_n)_{n \geq 1}$ with $\liminf(B_n) = \Z^2$.   
If  $(A_n)_{n \geq 1}$ is a $\RR$-valued sequence with
 $\liminf (  A_n) =\Z^2$,
then
\bea
|A_n|^{-1} H(X;A_n)  \toL \E Y_\0  {\rm ~~~~as~~} n \to \infty.    
\label{genln}
\eea
\end{lemm}

Let $\F_\0$ be the $\sigma$-field generated by
$(X_y,y\preceq  \0)$, where $y \preceq \0$ means $y$ precedes or equals $\0$
in the lexicographic ordering on $\Z^2$.
The following general CLT is a corollary of
Theorem  2.1 of \cite{P} (see Remark (iii) thereafter in \cite{P}).

\begin{lemm}
\label{genclt} 
Suppose $(H(X;A); A \in \RR) $  is a stationary $\RR$-indexed
functional of $X$ which stabilizes on sequences tending to $\Z^2$, 
and for some $\gamma >2$ satisfies 
\begin{equation}
\sup_{A \in \RR} \E[|\Delta_0(A) |^\gamma  ] < \infty.
\label{4moments}
\end{equation}
  Suppose that
$(A_n)_{n \geq 1}$ is a $\RR$-valued sequence with
$\liminf(A_n)= \Z^d$.  Then as $n \to \infty$,
$
|A_n|^{-1} \var(H(X;A_n)) $
converges to $ \sigma^2$,
and
\begin{equation}
|A_n|^{-1/2} ( H(X;A_n) - \E H(X;A_n) ) \tod \NN(0, \sigma^2),
\label{limdist}
\end{equation}
with $\sigma^2 = \E[( \E[\Delta_0(\infty)|\F_0])^2 ]$ .
\end{lemm}

We also use the following CLT for stationary random fields,
 from Bolthausen \cite{Bolt}.  For $A_1,A_2 \subset \Z^2$, let 
$ d(A_1,A_2)=  \inf\{(\|z_1-z_2\|: z_i \in A_i, i=1,2\}.$
Let $\partial A_1$ be the set of $z \in \Z^2 \setminus A_1$ such that
$d(A_1,\{z\})=1$.

\begin{lemm}
\label{lembolt} \cite{Bolt}
Suppose $(\psi_x, x \in \Z^2)$ is a 
real-valued stationary random field.
For integers $a_1,a_2,n \geq 1$ define
\bean
\alpha_{a_1,a_2}(n) = \sup \{ |P[F_1\cap F_2] -P[F_1]P[F_2]|:
F_i \in \sigma(\psi_z: z \in A_i), \\
 |A_i| \leq a_i, d(A_1,A_2) \geq n \}.  
\eean
Suppose $\sum_{m=1}^\infty m \alpha_{a_1,a_2}(m) < \infty $
for $a_1 + a_2 \leq 4$, and $\alpha_{1,\infty}(m)= o(m^{-2})$,
and $E[|\psi_\0|^3] < \infty$, and
$ \sum_{m = 1}^\infty m \alpha_{1,1}(m)^{1/3} < \infty$.

Then  $ \tsigma^2 := \sum_{z \in \Z^2} {\rm Cov}(\psi_\0,\psi_z) $
converges absolutely, and if $\tsigma^2 >0$,
then for any sequence $(\Gamma_n)_{n \geq 1}$
of subsets of $\Z^2$ with
 $|\partial (\Gamma_n)|/|\Gamma_n|\to 0$ as $n \to \infty$,
$ |\Gamma_n|^{-1/2} \sum_{z \in \Gamma_n} \psi_z \tod \NN(0,\tsigma^2)$. 
\end{lemm}

\section{Proof of LLN}
\label{secllnpf}
\allco

Let $\Po$ be a homogeneous Poisson point process 
of unit intensity on $\R^2 \times [0,\infty)$.  Given  $A \in \RR$, 
 label the points of the restriction of $\Po$ to
$\tilde{A} \times [0,\infty)$ as $\{(X_i(A),T_i(A))\}_{i=1}^{\infty}$ 
with  $T_1(A) <T_2(A) <T_3(A) < \cdots$.
Throughout the proofs of Theorems \ref{th2}-\ref{th2c},
we assume without loss of generality
 that the random input for the variable $N(A)$, 
defined in the introduction,  is given by the 
sequence of variables
$X_1(A),X_2(A),\ldots$ representing the 
locations of successive incoming particles (thus $T_i(A)$ is
taken to be the time of arrival of the $i$-th incoming particle).
By this device, coupled realizations of
 $N(A)$ are defined for all $A \in \RR$ 
simultaneously.

For each point $(X,T)$ of the restriction of
$\Po$ to $\tilde{A} \times [0,\infty)$,
define the pair $I(X,T;A)=(I_0(X,T;A),I_\rightarrow(X,T;A))$,
with $I_0(X,T;A)\in \{0,1\}$ and
$I_\rightarrow(X,T;A) \in \R^2$, as follows.
Let $I_0(X,T;A) $ (an indicator variable)
 be equal to 1 if the ball arriving
 at location $X$ at time $T$ 
is accepted, and to zero if it is rejected, in
the realisation of the BD model  with target set $\tilde{A}$
described above.
If $I_0(X,T;A) =1 $, let $I_\rightarrow(X,T;A)$ 
 denote the lateral displacement received by
the particle arriving at $X$ at time $T$,
prior to being adsorbed.
If $I_0(X,T;A) =0 $, 
 let $I_\rightarrow(X,T;A) =\0$. 


  By Lemma \ref{tennislem}, the decision on whether to 
accept an incoming particle, and also its displacement if accepted, 
are determined by the locations (after displacement) 
of those particles lying within a 
distance at most $10 \rho$ from the location
at which the new ball arrives.

The proof of the LLN and CLT involves a graphical construction.
  Make  the Poisson process $\Po$ on $\R^2 \times [0,\infty)$
 into the vertex set of an (infinite) oriented graph, denoted  $\G$,
by putting in an oriented  edge  $(X,T) \to (X',T')$ whenever
$\|X' - X \| \leq  20 \rho$  and $T < T'$. 
Observe that particle $(X,T)$ cannot affect
$(X',T')$ directly unless there is an edge $(X,T)\to(X',T')$
or the toroidal  boundary  conditions come into play.

For $z = (z_1,z_2) \in \Z^2$, and $\eps >0$, define
the squares
$$
Q_{z,\eps} :=  ((z_1-1)\eps, z_1 \eps]\times((z_2- 1)\eps, z_2 \eps];
~~~~ Q_z := Q_{z,1}.
$$  

For $x,y \in \Z^2$, let us say that {\em  $y$ is affected by $x$ before
 time $t$} if there exists a (directed) path  in
 the oriented graph  that starts at some Poisson point $(X,T)$ with
$X \in Q_x$, and ends at some  Poisson point $(Y,U)$ with
$Y \in Q_y$ and $U \leq t$.
Let $E_t(x,y)$
denote the event that
 $y $ is affected by $x$ before time $t$.

\begin{lemm}
\label{fpplem3}
There is a   constant  $\delta_1 \in (0,1)$  such 
that for  all $x,y \in \Z^d$,
$$
P[E_{\delta_1 \|x-y\|}(x,y)]  \leq 2 (3^{-\|x-y\|}).
$$
\end{lemm}
{\em Proof.} See Lemma 3.1 of  \cite{Plln}. This applies
directly if $20  \rho \leq 1$, and its proof  is easily adapted to the case 
$20  \rho > 1$.
$\qed$  \\


For $z \in \Z^d$ and $t >0$,  define the {\em cluster} $C_{z,t} \subset
\Z^d$ by
\bea
C_{z,t} :=  \{x \in \Z^d:  z {\rm ~is~affected~by~}x{\rm ~ before~ time ~}t\},
\label{clusdef}
\eea 
which is almost surely finite by Lemma \ref{fpplem3} and the Borel-Cantelli
lemma.
Let
$ B_{z,t} $ be the smallest element of $\RR$ that contains
$\cup_{x \in C_{z,t}} B(x;4 + 20 \rho)$. Note that
$B_{z,t}$ includes a `buffer zone' around $C_{z,t}$ so that
$$
{\rm dist} \left(  \cup_{y \in \Z^d \setminus B_{z,t}} Q_y  ,
 \cup_{x \in C_{z,t} } Q_{x} \right)  
> 20\rho,
$$
so that even if we were to add extra points
outside the union of squares $Q_y$,   $y \in B_{z,t} $,
there will not be any connected
 path in the graph from any
  of these added points to
  any Poisson point $(X,T) \in Q_z  \times (0,t]$.
This will be important later on.
\begin{lemm}
\label{lemar22}
Suppose $z \in \Z^2, t >0$. If $A$ is a lattice box
with $ B_{z,t} \subseteq  A $, then  
for all Poisson points $(X,T)$ lying in
$ Q_z \times [0,t]$ 
we have $I(X,T;A) = I(X,T;B_{z,t})$.
\end{lemm}
{\em Proof.}
By definition,
the influence of Poisson points outside $B_{z,t}$ does not propagate
to any Poisson points in
$ Q_z \times [0,t]$.  Therefore the fate of such points is the
same whether the target region is $\tilde{A}$ or $\tilde{B}_{z,t}$.
$\qed$ \\

Let $\S_0$ be the space of all finite subsets
$S$ of $D(\0;10 \rho)$ such that $\|x-y\|\geq 2\rho $
for all distinct $x,y\in S$.
Define $\Psi_0: \S_0 \to \{0,1\}$ and 
$\Psi_\rightarrow: \S_0 \to \R^2$ as follows.
For $S \in \S_0$, let $\Psi_0(S)$  take the value 1 (respectively 0)
if an incoming particle at the origin of $\R^2$ is accepted
(respectively rejected), given that $S$ is the configuration of 
previously accepted  particles in $B(\0;10 \rho)$.
If $\Psi_0(S)=1$, let $\Psi_{\rightarrow}(S) \in \R^2$  be the (lateral)
displacement of an incoming particle
 at the origin of $\R^2$, prior to acceptance,
given that $\S$ is the configuration of 
previously accepted  particles in $B(\0;10 \rho)$.
If  $\Psi_0(S)=0$, let $\Psi_{\rightarrow}(S) =\0 \in \R^2$.

We construct a spatially homogeneous form of the BD model with the
 whole of $\R^2$ as target region, as follows.  Define subsets $G_i$,
$i=0,1,2,3,\ldots$, of $\Po$ as follows.  Let $G_0$ be the set of roots of $\G$,
and recursively, if $G_0, \ldots, G_{n}$ are defined, let $G_{n+1}$ be
 the set of roots of the graph $\G$ with all vertices
in $G_0,\ldots,G_n$ removed.  As a consequence of
Lemma \ref{fpplem3}, 
the sets $G_0, G_1, G_2,\ldots$ form a partition of $ \Po$ (see 
 \cite{Plln}, Lemma 3.2).  

 Define  $I(X,T) = (I_0(X,T),I_\rightarrow(X,T) )$ with
$I_0(X,T) \in \{ 0,1\}$ and $I_\rightarrow(X,T) \in \R^2$
(representing acceptance status and lateral  displacement respectively)
for $(X,T) \in G_0,G_1,\ldots$ as follows.
If $(X,T) \in G_0$ 
then set $I_0(X,T) =1$ and 
 $I_\rightarrow (X,T) = \0$. 
Recursively for $n=1,2,3,\ldots$, 
for $(X,T) \in G_n$,  set
\bean
S_{X,T} = \{ Y + I_\rightarrow(Y,U) -X: (Y,U) \in \cup_{m=0}^{n-1}G_{m}, 
I_0(Y,U) = 1, 
\\
\|Y+ I_\rightarrow(Y,U) - X \| \leq 10 \rho  \},
\eean
then set $I_0(X,T) = \Psi_0 ( S_{X,T})$ and
$I_\rightarrow(X,T) = \Psi_\rightarrow (S_{X,T})$.

%
 For $t >0$, let $\xi_t$ be the point process of
  positions {\em after displacement} of particles accepted up to time $t$;
 that is, re-labelling the points
 of $\Po$ in arbitrary order as $\{(X_j,T_j) \}_{j=1}^\infty$,
 let $\xi_t$ be  the random locally finite set in $\R^2$ defined
 by
 \bea
 \xi_t = \{X_j + I_\rightarrow(X_j,T_j): I_0(X_j,T_j) = 1, T_j \leq t \}.
\label{xitdef}
 \eea
This point process is  now rigorously defined in terms of the
 Poisson process $\Po$ and the graphical construction.
It is a stationary point process on $\R^2$.
Define the limiting point process 
\bea
\xi = \cup_{t \geq 0} \xi_t.
\label{xidef}
\eea

Similarly, for $A \in \RR$ let  $\xi_t^A $ be the set of
locations after displacement (using the toroidal boundary conditions)  
of points $(X,T)$ of $\Po \cap ( \tilde{A} \times [0,\infty) )$
such that $T \leq t$ and $I_0(X,T;A) =1$.
All points of $\xi_t^A$ lie in $\tA$.
Define the limiting  point process
 $\xi^A = \cup_{t \geq 0} \xi_t^A$, 
a point process in $\tA$.

Choose $\eps_2 \in (0, \eps_0/2)$, with
 $\eps_0$ taken from Lemma \ref{bdlem1},
 and with $1/\eps_2 \in \NNN$.
For $z \in \Z^d$,
let $\beta_{z}$ denote the (random) time at which the square $Q_{z,\eps_2}$ 
becomes blocked, i.e.
the first time at which the point process 
$\xi_t$ leaves no
part of the surface of the square $Q_{z,\eps_2}$  still available
 to adsorb a sphere.
For  $A \in \RR$,
define $\beta_{z}^A$
in the same way with respect to the point process $\xi_t^A$.
The next result says that the variables $\beta_z$ and $\beta_z^A$ 
 are almost surely finite and in fact their distributions
have  tails which decay exponentially, uniformly in
$z,A$.

\begin{lemm}
\label{bdcoro}
It is the case that
\bea
\limsup_{t \to \infty} 
 t^{-1}
\sup_{z \in \Z^2}
\{
 \log P[\beta_z \geq t] \} 
< 0,
\label{bdeq0}
\eea
and
\bea
\limsup_{t \to \infty} 
t^{-1}
\sup
\{ \log
P[\beta_z^A \geq t]: 
A \in \RR, z \in \Z^2, Q_{z,\eps_2} \subset \tilde{A} \}
< 0.
\label{bdeq1}
\eea
\end{lemm}
{\em Proof.}
Suppose $\beta_z>t$, i.e.
 at time $t$ there exists a point $x \in Q_{z,\eps_2}$ that
is not yet covered, i.e. still available. Then by Lemma \ref{bdlem1},
 the probability 
of a particle arriving and causing  $Q_{z,\eps_2}$
to be covered by time $t+h$ is at least 
$\eps_1 h + o(h)$. Therefore we can choose $h_1 >0$
such that
$$
P[\beta_z> {t+ h_1 } |\beta_z >t] \leq
 1 - \eps_1 h_1/2,
$$ 
so that by induction, $P[\beta_z >nh_1 ]\leq
 (1 - \eps_1 h_1/2)^n $ for all $n \in \NNN$.
This argument holds uniformly in $z$,
and (\ref{bdeq0}) follows. 
Furthermore, the same argument  carries through to 
 the torus, to give (\ref{bdeq1}).
$ \qed$ \\

For $y \in \Z^d$, let
\bea
J_y = \max \{  \beta_{z}: Q_{z,\eps_2} \subset Q_x, x \in B(y;4+ 20\rho)\},   
\label{Jdef}
\eea
and (for $y \in A \in \RR$, with  $\| \cdot\|_A$ denoting the toroidal
metric)
\bea
J^A_y = \max \{  \beta^A_{z}: Q_{z,\eps_2} \subset Q_x, x \in A, \|x-y\|_A 
\leq 4+ 20\rho\}.   
\label{Jdef2}
\eea

For each $x\in B(y; 4+20\rho)$, the square $Q_x$ is  jammed
by the point process  $\xi_{J_y}$, 
 meaning that it is not possible for any Poisson point
 arriving  
 after  time $J_y$ to be accepted at a position in $Q_x$.
In particular, by Lemma \ref{tennislem}, all particles arriving
in $Q_y$ after time $J_y$ are rejected.
Define the enlarged `cluster' $B'_y$ by
%
$$
B'_y = \cup_{x \in B(y;4+20  \rho) } B_{x,J_y}.
$$
Using this enlarged cluster we can strengthen Lemma \ref{lemar22}
to account for arrivals at {\em all} times, as follows.
\begin{lemm}
\label{lemar22a}
Suppose  $y \in \Z^2$. If $A$ is a lattice rectangle
with $ B'_{y} \subseteq  A $, then  
for all Poisson points $(X,T)$ lying in
$ Q_y \times [0,\infty)$ 
we have $I(X,T;A) = I(X,T;B'_{y}) = I(X,T).$ 
\end{lemm}
{\em Proof.}
Suppose $(X,T)$ is a Poisson point in $Q_x \times [0,J_y]$,
for some $x \in B(y;4 + 20 \rho)$.
Then by
  Lemma \ref{lemar22}, we have 
$I(X,T;A) = I(X,T;B'_y) = I(X,T)$.

By definition of $J_y$,
it follows that  the restriction of the point
process $\xi_{J_y}^A$ to the set $\cup_{x \in B(y;2 + 10 \rho)}Q_x$
precludes the subsequent adsorption
of any more particles in
$\cup_{x \in B(y;2 + 10 \rho)}Q_x$, and in particular prevents
acceptance of any subsequent particles
arriving in $Q_y$;  therefore
for every  Poisson point  $ (X,T) \in Q_y \times (J_y,\infty)$,
we have
$I(X,T;A) = I(X,T;B'_y) =(0,\0)$.
$\qed$ \\

By Lemma  \ref{bdcoro}, the variable $J_y$ has an exponentially
decaying tail, uniformly in $y$, i.e.
\bea
\limsup_{t \to \infty} t^{-1} \sup_{y \in \Z^d} \log P[J_y > t] < 0.
\label{eq23a}
\eea
For $z \in \Z^d$,
let $\X_z$ be the image of the restriction of $\Po$ to $Q_z \times
 [0,\infty)$, under the translation that sends each point
$(X,T)$ to $(X-z,t)$. This is a homogeneous Poisson
point process on $Q_\0 \times [0,\infty)$.
The  Poisson processes $(\X_z, z \in \Z^2)$
are independent identically distributed random elements
of a measurable space $(E, {\cal E})$, where $E$ is the space
of locally finite subsets of $Q_\0 \times  [0,\infty) $.
The idea behind the proof of  Theorems \ref{th2} 
and \ref{clt2} is to regard
$N(A)$
as a stationary $\RR$-indexed functional 
driven by the process $\X=(\X_z)_{z\in \Z^2}$,
and  use Lemmas \ref{genllnclt} and \ref{genclt}
 from Section \ref{secprob}.
We write $\X$ rather than $X$ in this case. \\

{\em Proof of Theorem \ref{th2}.}
For $A \in \RR$ and $z \in \Z^2$, 
set
\bea
Y_{z}(\X;A) = 
\sum_{(X,T) \in \Po \cap (Q_z \times [0,\infty)) } I(X,T;A). 
\label{Ybd}
\eea
Then $(Y_z(\X;A), A \in \RR, z \in A)$
defined by (\ref{Ybd})
 is  a stationary $\RR$-indexed summand on
 on the process $\X = (\X_z)_{z \in \Z^2}$,
and  the corresponding 
stationary $\RR$-indexed sum $H(\X;A)$ is equal 
to $N(A)$. It suffices to check the 
  conditions in the general result Lemma \ref{genllnclt}.
Since the variables $Y_z(\X;A)$ are uniformly bounded
by a constant, (\ref{unimom}) holds.

Suppose $(A_n)_{n \geq 1}$ is a $\RR$-valued sequence
with $\liminf (A_n)= \Z^d$.
Then there exists a random variable $N_1$ such
 that for all $n \geq N_1$,
 $B'_{\0} \subseteq A_n$.
 By Lemma \ref{lemar22a}, for all $n \geq N_1$ and
 all Poisson points in
$Q_\0 \times [0,\infty)$ we have $I(X,T; A_n) = I(X,T;B'_{\0})$.  
%
Hence 
 $Y_\0(\X;A_n) = Y_\0(\X;B'_{\0})$ for $n \geq N_1$, so
 $Y_\0(\X;A_n)$ converges almost surely to
$Y_\0(\X;B'_{\0})$ as $n \to \infty$. Therefore 
  Lemma \ref{genllnclt} applies 
to give us (\ref{eqth2}) with convergence in $L^1$.
Convergence in $L^p$ then follows since
$N(A)/|A|$ is uniformly bounded by a constant. $\qed$ \\

{\em Proof of Theorem \ref{th2c}.}
Suppose $(A_n)_{n \geq 1}$ is a $\RR$-valued sequence 
with $\liminf( A_n ) = \Z^2$.
 It suffices to prove that for
any  bounded Borel $B \subset \R^d$ we have {\em almost sure} 
convergence 
\bea
\xi^{A_n}(B) \to \xi (B). 
\label{coup1}
\eea

 If $n$ is sufficiently large so that $B'_{z} \subseteq A_n$
for all $z$ within distance $4+10\rho$ of $B$,
then by   Lemma \ref{lemar22a},  
$
\xi^{A_n}(B) = 
\xi(B),
$
which  gives us  (\ref{coup1}). $\qed$

\section{Proof of CLTs}
\allco

With $\X$ as defined in the previous section,
$(N(B), B\in \RR)$ is a  
 stationary $\RR$-indexed functional driven
by the white noise process $\X$.
Our goal is to apply Lemma \ref{genclt}.

In this setting, the process $\X'$ appearing in the conditions
for  Lemma \ref{genclt}
is obtained from the process $\Po'$ defined as follows.
Let $\X'$ be the  Poisson process  obtained by
replacing the restriction   $\X_\0$ of 
$\Po$ to $Q_\0 \times [0,\infty)$
 with an   independent Poisson process $\X_*$ on
$Q_\0 \times [0,\infty)$, so that 
\bea
\X' = (\Po \setminus (Q_\0 \times [0,\infty) ) ) \cup \X_*. 
\label{po2def}
\eea

The points of $\Po' \setminus (Q_\0 \times [0,\infty))$
 are the same as those of $\Po \setminus (Q_\0 \times [0,\infty))$.
 However, the decision on whether to accept may be different; 
let $I'(X,T)=(I'_0(X,T),I'_\rightarrow(X,T))$ 
be defined in the same manner as $I(X,T)$ but
based on the process
 generated by $\Po'$ rather than $\Po$; 
likewise, given $A \in {\cal B}$,
 for $(X,T)\in \Po'\cap( \tA \times [0,\infty))$ 
let $I'(X,T;A) =(I'_0(X,T;A),I'_\rightarrow(X,T;A))$ 
 be defined in the same manner as $I(X,T;A)$ but 
based on the process generated by $\Po'$ rather than $\Po$. 

\begin{lemm}
\label{lem7a1}
 Let $z \in \Z^d$, and $t >0$.
Suppose $\0 \notin B'_{z}$.
 Then $I(X,T;A) = I'(X,T;A)$
for all points $(X,T)$ of $\Po$
in $ Q_z \times [0,\infty)$,
and all $A \in \RR$
with $B'_{z} \subset A$.
\end{lemm}
{\em Proof.}
Suppose $(X,T)$ is a point  of $\Po$, with $X \in Q_z$.
Then
$$
I(X,T;A) = I(X,T;B'_{z}) = I'(X,T;B'_{z}) = I'(X,T;A), 
$$
where the first equality comes from Lemma \ref{lemar22a},
and the second comes from the equality  of $\Po$ and 
$\Po'$ outside $Q_\0\times[0,\infty)$.
\qed \\

The idea for proving stabilization goes as follows. By Lemma \ref{fpplem3},
the effect of changing the inputs at the origin propagates like an
`infection' spreading through space at a 
linear rate. However, if this `infection' encounters a `wall' of 
thickness $10\rho$ surrounding the origin, consisting of sites which are
entirely blocked before the infection reaches them, then
this wall prevents any spread of the effect of changing inputs at
the origin to the other side of the wall.
 The existence with
high probability of
a wall surrounding the origin follows from 
the fact that the probability that a  site is not blocked
by time $t$ decays exponentially in $t$. 

By Lemma \ref{fpplem3} and (\ref{eq23a}),
$P[\0 \in B'_{z} ]$
decays exponentially in $\|z\|$. 
Therefore,  by the Borel-Cantelli lemma,
\bea
P[ \0 \in B'_{z} \mbox{ for infinitely many } z] =0.
\label{apr5}
\eea


For $t>0$, define the  annular region
$$
\A_t :=   \cup_{z \in B(\0;t + 4 + 20 \rho  ) \setminus B(\0;  t)}
Q_z
$$
and the `distant' set 
$$
\D_t : =  \cup_{z \in \Z^2 \setminus B(\0; t + 4 + 20 \rho )} Q_z.
$$
\begin{lemm}
\label{lem7a}
Let $t >0$, and
suppose $\0 \notin B'_{z} $ for all 
$  z \in \A_t \cap \Z^d$. Then  
 $I(X,T;A) = I'(X,T;A)$
for all points $(X,T)$ of $\Po$
in $ \A_t \cup \D_t$,
and all 
$A \in \RR$ with $B'_{z} \subset A$ for all $z \in {\cal A}_t \cap \Z^d$.
\end{lemm}
{\em Proof.}
 Suppose $\0 \notin B'_{z} $ for all 
 $  z \in \A_t$.  Suppose also
$A \in \RR$ with $B'_{z} \subset A$ for all $z \in {\cal A}_t$.
By Lemma \ref{lem7a1},
\bea
I(X,T;A)  = I'(X,T;A), ~~~~ \forall (X,T)  \in \Po \cap 
( {\cal A}_t \times [0,\infty)) .
\label{eq7b}
\eea

Next consider
Poisson points $(X,T)$ with $X \in \D_t$. 
Any path to $(X,T)$ from $Q_\0 \times [0,\infty)$
must pass through 
 $ \A_t \times  [0,\infty)$,
and the status of all points in this region is unaffected
by the change in $\Po_{\0}$, so that
$I(X,T;A) = I'(X,T;A)$. More formally, we use an induction, as follows.

Define generations $G_0(A,t), G_1(A,t), \ldots$ as follows.
Let $\G(A,t)$ be
the restriction of $\G$ to vertices in $(\D_t \cap \tA) \times [0,\infty)$.
Let $G_0(A,t) $ be the set of roots  of $\G(A,t)$. 
Then remove
vertices of $G_0(A,t)$ from   
 $\G(A,t)$ and let $G_1(A,t)$
 be the set of roots of the remaining
oriented graph. Then remove vertices in $G_1(A,t)$ too and
let $G_2(A,t)$ be the set of roots of the remaining graph and so on.
The sets $G_0(A,t), G_1(A,t), \ldots$ form a partition of 
the vertex set of $\G(A,t)$, because
 $C_{z,t}$ defined at (\ref{clusdef}) is 
finite for all $z,t$.

The inductive hypothesis is  that  
if $(X,T) \in   G_n(A,t)$, then $I(X,T;A) = I'(X,T;A) $.
   This is true for $n=0$, because if $(X,T) \in G_0(A,t)$, 
then  any $(X',T')$ for which $X'\in \tA$ and there is an edge
from $(X',T')$ to $(X,T) $ lies
%
%
in the annulus
$\A_t \times [0,\infty)$,  and therefore by (\ref{eq7b}) satisfies 
$I'(X',T';A) = I(X',T';A)$, which implies   
$I'(X,T;A) = I(X,T;A)$,
since the decision on the value of $I(X,T;A)$ 
depends only on the   
decisions at points $(X',T')$ from which there are edges
to $(X,T)$

Now  suppose the hypothesis is true for $n= 0,1,\ldots, k-1$.
Then if $(X,T) \in G_k(A,t)$ 
all of the points $(X',T')$ from which there is an edge to
$(X,T)$ 
lie either  in one of the  generations $G_0(A,t), \ldots, 
G_{k-1}(A,t) $, or in
$\A_t \times [0,\infty)$,  and therefore,  by the inductive
hypothesis and
by (\ref{eq7b}),  all such $(z',T')$   satisfy 
$I(z',T';A) = I'(z',T';A)$, and hence again
$I'(z,T;A) = I(z,T;A)$.
$\qed$ \\

\begin{lemm}
\label{checkmoments}
If for $A \in \RR$ we set 
$$
\Delta_0(A) = \sum_{z \in A} \sum_{(X,T) \in \Po: X \in Q_z }(I(X,T;A) - I'(X,T;A)) 
$$ 
then
\bea
\sup_{A \in \RR} \E[|\Delta_0(A) |^3  ] < \infty.
\label{4moments2}
\eea
\end{lemm}
{\em Proof.}
 Modify  the 
 graphical construction from Section \ref{secllnpf} 
to produce an oriented graph $\G^A$ 
with vertex set the Poisson process $\Po \cap (\tA \times [0,\infty) ) $,
by putting in an oriented  edge  $(X,T) \to (X',T')$ whenever
$\|X' - X \|_A \leq  20 \rho$  and $T < T'$, where
$\|\cdot \|_A$ denotes the toroidal metric. 
For $x,y \in A$, let us say that {\em  $y$ is affected in $A$ by $x$ before
 time $t$} if there exists a (directed) path  in
 the oriented graph $\G_A$  that starts at some Poisson point $(X,T)$ with
$X \in Q_x$, and ends at some  Poisson point $(Y,U)$ with
$Y \in Q_y$ and $U \leq t$.
For $z \in A$ and $t >0$,  define the `$A$-cluster' $C^A_{z}$
to consist of all
$x \in A$ such that some $y \in A$
with  $\|y-z\|_A\leq 4 + 20 \rho$
 is affected in $A$  by $x$ before time 
$J_y^A $.

A similar argument to the proof of Lemma \ref{lem7a1} yields
 $I(X,T;A) = I'(X,T;A)$
for all $(X,T)$ with $X \in Q_z$ and 
$B(\0;4 + 20\rho) \cap C^A_{z} = \emptyset$.
Hence for any $A \in \RR$,
$$
 \Delta_\0 (A) \leq c 
\sum_{z \in A}  {\bf 1}_{ \{B(\0;4 + 20 \rho) \cap C^A_{z}  \neq \emptyset
 \}}.
$$
However, $P[B(\0;4 + 20 \rho) \cap  C^A_{z} \neq \emptyset]$ 
 decays exponentially in $\|z\|$, uniformly over $A\in \RR$,
 because  of Lemmas \ref{fpplem3} and 
\ref{bdcoro}.  The bounded moments 
condition (\ref{4moments2}) follows.
$\qed$ \\

 {\em Proof of Theorem \ref{clt2}.}
As in the proof of Theorem \ref{th2}, 
define $Y_z(\X;A)$ by (\ref{Ybd}). 
The aim is to apply
 Lemma \ref{genclt}.
The bounded moments condition (\ref{4moments})
follows from Lemma \ref{checkmoments}.

We need 
to check  stabilization. By (\ref{apr5}), there exists
a random $R$ such that $0 \notin B'_z$  for 
$z \in \Z^d $ with $\|z \| \geq  R $.  Let
$B_\infty$ be the smallest element of $\B$ containing
$\cup_{z \in B(\0;z)} B'_z$.
Suppose $(A_n)_{n \geq 1}$ is a $\RR$-valued sequence
with $\liminf(A_n)= \Z^d$.  
Then there exists random $N_2$ such that 
$$
B_\infty
\subset A_n, \mbox{ for all }
n \geq N_2.
$$
and such that a similar expression holds with regard
to the Poisson process $\Po'$ rather than $\Po$.
Then by Lemma \ref{lemar22a}, 
 for all $n \geq N_2$ and for all
 $z \in  B(\0; R)$, we have
\bea
Y_z(\X;A_n) - Y_z(\X';A_n) = 
Y_z(\X;B_{\infty}) - Y_z(\X';B_{\infty})
\mbox{ for all } n \geq N_2.
\label{N2}
\eea
Also, there exists random $N_3 \geq N_2$ such that for all 
$n \geq N_3$
we have 
$$
B'_z \subset A_n \mbox{ for all } z \in {\cal A}_{R} \cap \Z^d.
$$
Hence by the definition of $R$ and Lemma \ref{lem7a},
for all $z \in \Z^d \setminus  B(\0; R)$,
$$
Y_z(\X;A_n) = Y_z(\X';A_n) \mbox{ for all } n \geq N_3.
$$
Combined with (\ref{N2}), this gives us
for all $ n \geq N_3$,
$$
\sum_{z \in A_n}
\left( Y_z(\X;A_n) - Y_z(\X';A_n) \right) = 
\sum_{z \in B(\0; R)}
(Y_z(\X;B_\infty) - Y_z(\X';B_\infty)),
\label{N3}
$$
which demonstrates  stabilization  of
the  induced functional 
$$
H(\X;A) = \sum_{y \in A} Y_y(\X;A) = N(A).
$$ 
Therefore all the conditions for Lemma \ref{genclt} hold here,
and by that result  the conclusion
of Theorem \ref{clt2} holds, subject to
showing that $\sigma_1 >0$.

The value of $\sigma_1$  is independent of the
choice of sequence $(A_n)$ (provided $\liminf (A_n) = \Z^2$)
  and therefore to show $\sigma_1>0$
using (\ref{varcon2}) we are at liberty to choose any sequence
$(A_n)_{n \geq 1}$. Let $K= \lceil 200 \rho\rceil$.
Take $A_n $ to be a lattice square of side $Kn$,
and divide $\tA_n$ into squares 
of side $K$, which we shall  refer to  as {\em blocks}.

Inside each block $S_i$ let $T_i$ be the annulus of
thickness $18 \rho$, consisting of points
at a distance more than $2 \rho$ but less than $20 \rho$ from the 
boundary of the block.
Also let $S^-_i$ be the interior region consisting
of points at a distance more than $20 \rho$ from the boundary of the block. 
 Let $I_i$  be the indicator
random variable of the event that before 
there are any arrivals at all in $S_i \setminus T_i$, a sequence
of ball centres  arrive in $T_i$  in such a way
that the corresponding particles are adsorbed without
rolling and cause a barrier between the interior
region $S^-_i$ and the complement of $S_i$, by making 
all points in $T_i$ unavailable.

The probability $P[I_i=1]$
is very small but not zero,
and  does not depend  on $i$ or $n$ since $K$ is fixed.
Let $N_n = \sum_{i=1}^{n^2} I_i$. 
  Then $\E[N_n]/|A_n|$ is  a non-zero constant.

Let $\F$ be the $\sigma$-field generated by
the value of $N_n$, along with the positions of the accepted
particles  not lying in the union of the   squares $\{S_i^-: I_i=1\}$.
Then
\bean
\Var(H(A_n)) = \var( \E [H(A_n)|\F])
+ \E[ \var(H(A_n|\F))] \geq    
 \E[ \var(H(A_n|\F))].    
\eean
Suppose we are given the value of $N_n$ and the
configuration of accepted items
outside   the squares $S_i^-, I_i=1$.
The
only remaining variability is from the number of accepted
items inside the inner squares $S^-_i$
contributing to $N_n$.

Let $S_i^*$ be the square consisting of points
in $S_i$ at a distance at least $22 \rho$ from $S_i$,
i.e. a slightly smaller square inside $S^-_i$.
We consider two possible lattice configurations
inside $S_i^*$.

Let $\{x_{i,1}, \ldots,x_{i,n_1}\}$
 be the restriction to $S_i^*$ of a
regular triangular lattice with
each point distant $(2.02)\rho$ from its neighbours (tight packing).
Let $\{y_{i,1}, \ldots, y_{i,n_2}\}$ be the restriction to $S_i^*$ of a
regular triangular lattice with
each point distant $3\rho$ from its neighbours (loose packing).
Let $E_i$ be the event that the first $n_1$ particles
in $S_i^-$ to arrive are  in the small disks
$D(x_i;0.01), 1\leq i \leq n_1$, and are in different disks.
Let $F_i$ be the event that the first $n_2$ particles
in $S_i^-$ to arrive are  in the disks
$D(y_i;0.01), 1\leq i \leq n_2$, and are in distinct disks.

Events $E_i$ and $F_i$ have probability bounded away from zero.
More particles will be packed into the 
square $S_i^-$ on event $E_i$ than on event $F_i$.
It follows that  there is a constant $c>0$ such that
 given $N_n =k$, 
$
\var(H(A_n)|\F) \geq  ck. 
$
Therefore $\Var(H(A_n)) \geq c \E[N_n]$,
and this divided by $|A_n|$ is bounded away from zero.
Hence $\sigma_1>0$ by (\ref{varcon2}). 
This completes the proof.
$\qed$ \\

Finally we shall prove Theorem \ref{cltII}.
The aim is to apply Lemma \ref{lembolt}.
Let the point processes 
$\xi_t$ and $\xi$ (the set of locations of adsorbed  particles at time
 $t$ and at time $\infty$, respectively)
  be as defined at
(\ref{xitdef}) and (\ref{xidef}).
The family of variables $(\xi(Q_z), z \in \Z^2)$ forms a stationary random 
field.
We need to show rapidly decaying correlations for this random field,
and do so  via Lemmas \ref{lemapr30} and \ref{lemapr24} below.
 The first of these (but apparently not 
 the second)  can  be proved using Theorem 4.20
from Chapter I of  Liggett \cite{Li}, but we take a different
approach which is closer to that used already.  

For $z \in \Z^d$, 
let $B''_z$ be the union of all sets $B'_y, y \in B(z;10 \rho +2)$.
Lemmas \ref{fpplem3} and \ref{bdcoro}  imply an exponentially decaying
tail for the diameter of the set $B'_\0$, and hence 
the distribution of the
 diameter of $B''_z$ also has an exponentially decaying tail, uniformly
in $z$, i.e.
\bea
\limsup_{r \to \infty} \sup_{z \in \Z^d} r^{-1} \log P[ 
B''_z \cap (\Z^2 \setminus B(z;r) ) \neq \emptyset ] < 0
\label{apr30}
\eea

\begin{lemm}
\label{lemapr30}
  Given any finite $\Gamma \subset \Z^2$, let ${\cal F}_\Gamma$ be the
 $\sigma$-field generated by $(\xi(Q_z), z \in \Gamma)$.
There exist positive finite constants $K', \delta_2$
 such that if $\Gamma,\Gamma'$ are  sets in $\Z^2$, both of cardinality at most
 4, 
 and the distance between them is $d(\Gamma,\Gamma')$, we have for all  events 
$ F \in {\cal F}_{\Gamma}$, $G \in {\cal F}_{\Gamma'}$,
$$
P [ F \cap G] -P[F]P[G]
\leq  K' \exp(- \delta_2 d(\Gamma,\Gamma') ).
$$
\end{lemm}
{\em Proof.}
Let $\Po'$ be an independent copy of the Poisson
process $\Po$.  
 Suppose $F\in \F_{\Gamma}$ and $G \in \F_{\Gamma'}$. 
Then $F= \{ (\xi(Q_z))_{z \in \Gamma} \in R \}$
 for some Borel $R \subset
\R^\Gamma$, and 
 $G = \{ \xi(Q_z)_{z \in \Gamma'} \in R' )$ for some Borel $R' \subset
\R^{\Gamma'}$.

Let $H_F$ be the set of points of $\R^2$ lying closer to $F$
than to $G$ and let
 $H_G = \R^2 \setminus H_F$.
Let $F^*$ be defined like $F$ but based on points
of  the Poisson process
$$
\Q :=
(\Po \cap (H_F \times [0,\infty) ) ) \cup ( \Po' \cap
  (H_G \times [0,\infty) ) ).
$$
That is, let $ \xi^{(1)}$ be defined in the same manner as $\xi$
(eqn. (\ref{xidef})) but in terms of the Poisson process
$\Q$ instead of $\Po$, and let  
 $F^* := \{ (\xi^{(1)}(Q_z))_{z \in \Gamma} \in R \}$.
Similarly, 
 let $ G^*$ be defined like $G$ based on points of
$$
\Q' := (\Po \cap (H_G \times [0,\infty) ) ) \cup ( \Po' \cap
  (H_F  \times [0,\infty) ) ),
$$
that is, let $ \xi^{(2)}$ be defined in the same manner as $\xi$
(eqn. (\ref{xidef})) but in terms of the Poisson process
$\Q'$ instead of $\Po$, and let  
 $G^* := \{ (\xi^{(2)}(Q_z))_{z \in \Gamma} \in R' \}$.

Then $F^*$ and $G^*$ are independent (since based on independent
Poisson processes $\Q,\Q'$),  and
$P[F^*] = P[F]$, and $P[G^*]= P[G]$. Therefore
$$
P[ F \cap G] -P[F]P[G] = P[F\cap G ]- P[ F^* \cap G^*]
$$ 
so that
$$
|P[ F \cap G] -P[F]P[G]| \leq P[ F \triangle F^*] + P[G \triangle G^*].
$$ 
By Lemma \ref{lemar22a},
 $F \triangle F^*$ does not occur if
$Q_y \subset H_F$ for all $ y \in B''_{x}$ and all $x \in \Gamma_1$.
Likewise, $G \triangle G^*$ does not occur if
$Q_y \subset H_F$ for all $ y \in B''_{x}$ and all $x \in \Gamma_2$.
By (\ref{apr30}),
 $P[F \triangle F^*]$ and $P[G \triangle G^*]$
  both decay exponentially in $d(\Gamma_1,\Gamma_2)$,
uniformly over finite $\Gamma_i \subset \Z^2$
of cardinality at most 4 and over $F \in \F_{\Gamma_1}$,
 $G \in \F_{\Gamma_2}$.
$\qed$

\begin{lemm}
\label{lemapr24}
Let $\F_0 =\F_{\{\0\}}$ and
let $\F_t$ be the $\sigma$-field generated by
the variables $\xi(Q_z), z \in \Z^d \setminus B(t)$.
Then
$\sup\{|P[F\cap G] - P[F]P[G]|: F \in \F_0, G \in \F_t\}$
decays exponentially in $t$.
\end{lemm} 
{\em Proof}.
Let $\Po'$ be an independent copy of the Poisson
process $\Po$.  
 Suppose $F\in \F_0$ and $G \in \F_t$. 
Let $F^*$ be defined like $F$ but based on points
of 
$$
(\Po \cap (D(\0;t/2) \times [0,\infty) ) ) \cup ( \Po' \cap
  (\R^2 \setminus D(\0;t/2) ) \times [0,\infty) )
$$
and let $ G^*$ be defined like $G$ but based on points of
$$
(\Po' \cap (D(\0;t/2) \times [0,\infty) ) ) \cup ( \Po \cap
  (\R^2 \setminus D(\0;t/2) ) \times [0,\infty) ).
$$
The precise  definition of $F^*$ and $G^*$ is analogous to
that used in the preceding proof.
Then $F^*$ and $G^*$ are independent and
$P[F^*] = P[F]$, and $P[G^*]= P[G]$. Therefore, as in the preceding proof,
$$
|P[ F \cap G] -P[F]P[G]| \leq P[ F \triangle F^*] + P[G \triangle G^*].
$$ 
By Lemma \ref{lemar22a},
 $F \triangle F^*$ does not occur if
$B''_{\0} \subset B(\0;t/4)$.
Hence by (\ref{apr30}),
 $P[F \triangle F^*]$  decays exponentially in $t$,
uniformly over $F \in \F_\0$.

By an extension to the proof of Lemma \ref{lem7a},
$G \triangle G^*$ does not occur if
$$
B (\0;3t/4) \cap B'_z = \emptyset ~~~\forall z \in {\cal A}_{t-2} \cap \Z^d
$$
and the probability of this last event decays exponentially in $t$
by (\ref{apr30}).
$\qed$ \\

{\em Proof of Theorem \ref{cltII}.}
By lemmas \ref{lemapr30} and \ref{lemapr24},
the result  follows by taking $\Gamma_n = B_n$,
in Lemma \ref{lembolt}, provided we have $\sigma_2 >0$. An elementary
 argument shows  that with $\psi_z= \xi(Q_z)$,
$$
|B_n|^{-1} \Var \left( \sum_{x \in B_n} \psi_x \right)
\to \sum_{z \in \Z^2} {\rm Cov}(\psi_\0,\psi_z) :=\sigma_2^2,
$$
and the left hand side of this expression can be shown bounded
away from zero  by a similar argument to that used at the end of  
the proof of Theorem \ref{clt2}. Therefore $\sigma_2 >0$. $\qed$

\end{document}